\documentclass[twocolumn,showpacs,aps,prl,superscriptaddress,showkeys,,amsmath,amssymb]{revtex4}

\usepackage{graphicx}
\usepackage{bm}

\begin{document}


\title{
Exchange distortion and spin Jahn-Teller effect for triangular and tetrahedral spin clusters of spin-1/2
}
\author{Kiyosi Terao}
\email{terk005@shinshu-u.ac.jp}
\affiliation{Faculty of Science, Shinshu University, Matsumoto 390-8621, Japan}
\author{Ikumi \surname{Honda}}
\altaffiliation[Present address:]{
Ceratech Japan Co., Ltd.\\
500 Shinonoi Okada, Nagano 381-2295, Japan
}
\affiliation{Graduate School of Science and Technology}

\begin{abstract}
We study the effects of magneto-elastic coupling on the degenerate ground state of the antiferromagnetic Heisenberg model on regular triangle and tetrahedron clusters of spin-1/2.
Both give very similar results.
Static distortion lifts the degeneracy of the ground state through the distance dependence of the exchange coupling.
On the contrary, quantum-mechanical or dynamical distortion does not.
The tetragonal distortion at the non-magnetic phase transition of spinels is discussed.
\end{abstract}


\pacs{ 75.10.Jm, 75.40.Gb , 75.80.+q}
\keywords{spin Jahn-Teller effect , exchange striction , quantum spin , vanadium spinel}
\maketitle

The highly degenerate antiferromagnetic (AF) ground state of the pyrochlore lattice is very interesting.
Reduction of the degeneracy through the distance dependence of exchange coupling was investigated before by Terao~\cite{tera,ter2} with classical spins.
Detailed investigations into the spin Jahn-Teller effect were given for tetramer models with classical spin and quantum spin-1/2 by Tchernyshyov {\it et al.}~\cite{tche,tch2}
and by Yamashita and Ueda~\cite{YamashitaUeda00}, respectively, on the basis of static distortion models.
Recently, quantum-mechanical treatment of distortion was given by the present authors~\cite{HondaTerao1,HondaTerao2}.

In the present paper, quantum-mechanical consideration is given to the effects of distortion upon the degenerate ground spin-states of the AF Heisenberg Hamiltonian of spin-1/2 on the regular triangle and tetrahedron clusters.

\begin{figure}[hbt]
\begin{center}
\includegraphics[width=5.6cm]{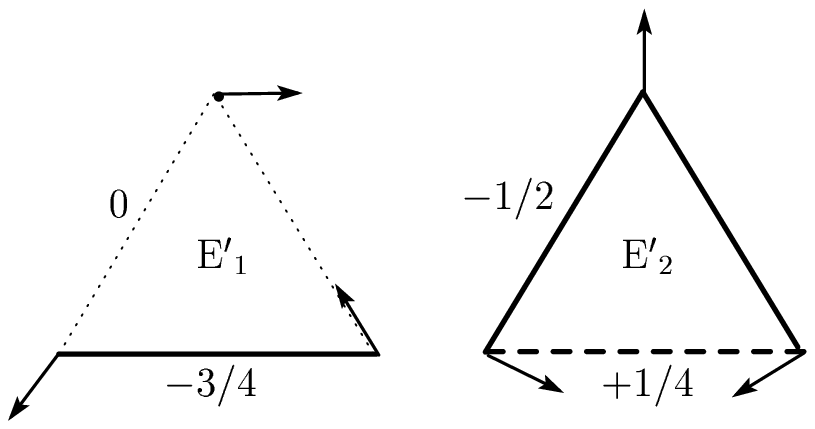}
\end{center}
\caption{Displacement (arrows) and spin correlations (numerals).}
		\label{fig:triangle}
\end{figure}

\begin{figure}[hbt]
	\begin{center}
		\includegraphics[width=7cm]{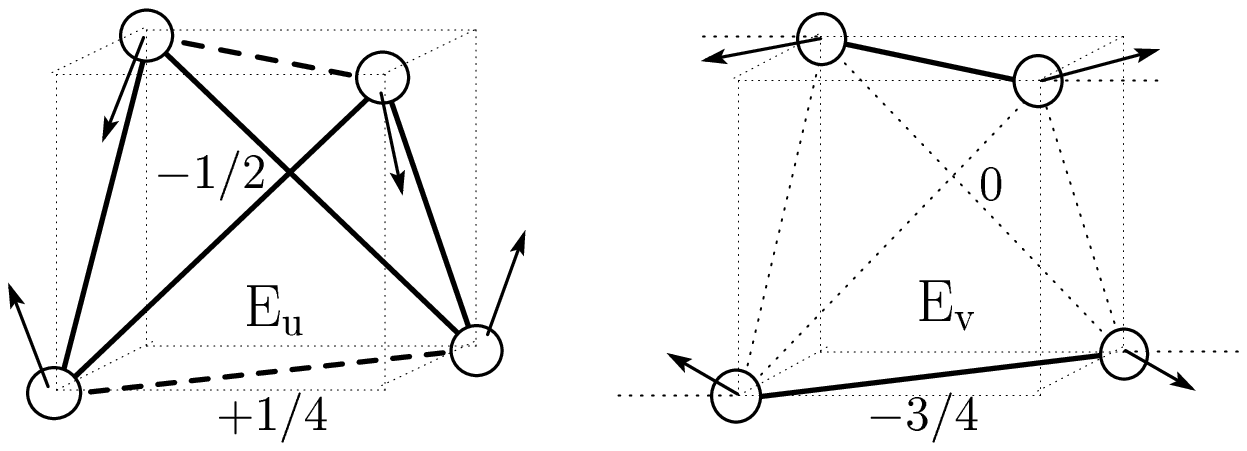}
	\end{center}
	\caption{Displacement (arrows) and spin correlations (numerals).}
		\label{fig:tetrahedron}
\end{figure}

The Hamiltonian for the regular cluster is
${\mathcal H}_0 \,=\, -2J_0 {\sum}_{\ell<\ell '} \bm{s}_\ell \cdot \bm{s}_{\ell '}$,
where $J_0 <$ 0 and $s$ = 1/2.
The AF ground states of the triangle and tetrahedron clusters conform to the doublet E$'$ and E representations for the D$_{3{\rm h}}$ and T$_{\rm d}$ point groups, respectively. 
The eigenvalue equations are 
$
	\mathcal{H}_0 |{\rm{E'}}i \!\!> \,=\, (3/2) J_0 |{\rm{E'}}i \!\!>, \   i=1,\, 2,
$
with the total spin $S$ = 1/2 for the triangle, and
$
	\mathcal{H}_0 |\,{\rm E} \eta \!\!> = 3J_0 |\,{\rm E} \eta \!\!>, \  \eta={\rm u,\, v},
$
with S = 0  for the tetrahedron.
We put aside the Kramers degeneracy.
The spin correlations ${<\!\!\bm{s}_\ell \cdot \bm{s}_{\ell '}\!\!>}$ in these states are shown in Figs.~\ref{fig:triangle} and \ref{fig:tetrahedron} by the numerals by the bonds.
The distortions for the triangle and the tetrahedron are, respectively, classified into $\rm{A_1'}$, $\rm{E'}$ ($Q_1,\ Q_2$, their normal coordinates) modes by D$_{3\rm h}$ group and into A$_1$, T$_2$, E ($Q_{\mathrm{u}},\, Q_{\mathrm{v}}$) modes by T$_{\rm d}$ group.
The E and E$'$ modes are illustrated in Figs.~\ref{fig:triangle} and \ref{fig:tetrahedron} by the arrows.
The perturbation Hamiltonians by the E and E$'$ modes for the triangle and the tetrahedron are
\begin{align}
\mathcal{H'}_{\rm E'} = &({P_1}^2+{P_2}^2)/2m
                  + m {\omega_{\rm E'}}^2( {Q_1}^2 + {Q_2}^2) /2 \nonumber\\
                  &- 2J'_{\rm E'}(Q_1 f_1+ Q_2 f_2) ,
			\label{eqkanaNU}
\end{align}
and
\begin{align}
	\mathcal{ H'}_{\rm E} =& ({P_{\rm u}}^2+{P_{\rm v}}^2)/2m 
	 + m {\omega_{\rm E}}^2 ( {Q_{\rm u}}^2 + {Q_{\rm v}}^2)  /2 \notag \\
	 &-2 J'_{\rm{E}}(Q_{\rm u} f_{\rm u}+ Q_{\rm v} f_{\rm v}),
			\label{eq:perturbation1}
\end{align}
where $f_{\alpha}$'s are  the bases for the irreducible representations made from $\bm{s}_\ell \cdot \bm{s}_{\ell '}$'s and $J'_\alpha $'s are the drivatives of $J$ with respect to distortion. 
For the E and E$'$ representations, 
$
	(f_1,\ f2) = ((\bm{s}_1\cdot\bm{s}_2-\bm{s}_3\cdot\bm{s}_1)/\sqrt2,\ 
	(\bm{s}_1\cdot\bm{s}_2-2\bm{s}_2\cdot\bm{s}_3+\bm{s}_3\cdot\bm{s}_1)/\sqrt6),
$
and
$
	(f_{\rm u},\ f_{\rm v}) = ( [(\bm{s}_1+\bm{s}_2)\cdot (\bm{s}_3+\bm{s}_4)
	-2(\bm{s}_1  \cdot \bm{s}_2 +\bm{s}_3 \cdot \bm{s}_4)] /\sqrt{12}, \ 
	 (\bm{s}_1 - \bm{s}_2) \cdot (\bm{s}_3 - \bm{s}_4)/2 ).
	\label{eq:fE}%
$
In the subspace of the ground spin-states, 
$  (f_1, \  f_2) = \sqrt6/4 \,(\sigma _x, \  \sigma_z)$
for the triangle and
$  (f_{\rm u}, \  f_{\rm v}) = -\sqrt3/2 \, ( \sigma _z,\ \sigma _x)$
for the tetrahedron, where $\sigma_z$ and $\sigma_x$ are the  Pauli matrices. Note that f$_\alpha$'s for the T$_2$ representation for the tetrahedron vanish and ${\rm T}_2 \times {\rm E}$ is reduced to ${\rm T}_1 + {\rm T}_2$, hence the T$_2$ representation terms are irrelevant. 

%
The static displacement splits the energies for the triangle and for the tetrahedron, respectively, as
\begin{align}
 \mp \sqrt{3/2} J'_{\rm E'}\sqrt{{Q_1}^2+{Q_2}^2}
\text{ and }
 \mp \sqrt{3}J'_{\rm{E}}\sqrt{{Q_{\rm u}}^2+{Q_{\rm v}}^2}.
\end{align}
Minimizing the energies, the changes in energies are estimated as
\begin{align}
\delta E' = - \frac{3 {J'_{\rm E'}}^2}{4 m {\omega_{\rm E'}}^2}
\text{\ \ at\ \ }
	\sqrt{{Q_1}^2+{Q_2}^2} =\sqrt{\frac{3}{2}}\frac{ |{J'_{\rm E'}|} }{ m {\omega_{\rm E'}}^2 }
			\label{eqkanaTE}%
\end{align}
for the triangle, and
\begin{align}
\delta E' = - \frac{ 3{J_{\rm E}'}^2 }{2 m {\omega_{\rm E}}^2 }
\text{\ \ at\ \ }
 	\sqrt{{Q_{\rm u}}^2+{Q_{\rm v}}^2}
	=\frac{\sqrt{3} |J'_{\rm E}| }{ m {\omega_{\rm E}}^2 }
			\label{eq:statQ}
\end{align}
for the tetrahedron.

We consider the dynamical displacement 
by representing $Q_\alpha $ by phonon operators, $b^\dagger_\alpha$ and $b_\alpha$.
The r.h.s. of Eqs.~(\ref{eqkanaNU}) and (\ref{eq:perturbation1}) are rewritten as
$
  \sum_\alpha [
    \hbar \omega _\alpha ( b_\alpha ^\dagger b _\alpha +1/2 \bigr)
    -\sqrt{  \hbar /2m\omega _\alpha } 
     {J'} _\alpha ( b_\alpha + b_\alpha ^\dagger ) f_\alpha 
     ].
$
We define modified operators
$
  \tilde{b}_\alpha = b_\alpha - \sqrt2 J'_\alpha f_\alpha/\sqrt{m\hbar \omega_\alpha^3},  
$
then
\begin{equation}
  \mathcal{H'} ={\sum}_{\alpha }
  [
    \hbar\omega_\alpha
    (
      \tilde{b}_\alpha^\dagger \tilde{b}_\alpha +1/2 
    )
  -2{J'_{\alpha }} ^2 f_{\alpha}^2 /m {\omega_\alpha }^2
  ].
             \label{eq:H1tild}
\end{equation}
Commutators of the modified operators are
$
[\tilde{b}_\alpha ,\tilde{b}_\alpha^\dagger ] =1
$
and
$ 
[\tilde b_\alpha ,\tilde b_\alpha ] =[\tilde b_\alpha^{\dagger} ,\tilde b_\alpha^{\dagger} ]=0,
$
so $\tilde b_\alpha$, $\tilde b_\alpha^\dagger$ are the Boson operators.
Although the excited states is complicated because of the commutation relations,
\begin{align}
	[\tilde b_1,\tilde b_2^{\dagger}]
   &= [\tilde b_1,\tilde b_2] = [\tilde b_1^{\dagger},\tilde b_2^{\dagger}] 
	=\frac{i 2\sqrt{3}{J'_{\rm E'}}^2 }{m\hbar{\omega_{\rm E'}}^3 } 
    \bm{s}_3 \cdot (\bm{s} _1 \times \bm{s} _2),\\
	[ \tilde{b}_{\rm u}, \tilde{b}_{\rm v} ^\dagger ] &= [\tilde{b}_{\rm u}, \tilde{b}_{\rm v}] 
	= [\tilde{b}_{\rm u}^\dagger, \tilde{b}_{\rm v}^\dagger]
	=  \frac{i\sqrt{3} {J'_{\rm E}}^2 }{ m \hbar \omega_{\rm E}^3} \notag \\
	\{ ( &\bf{s}_1 - \bf{s}_2) \cdot ( \bf{s}_3 \times \bf{s}_4) 
	+( \bf{s}_3 - \bf{s}_4) \cdot ( \bf{s}_1 \times \bf{s}_2) \},
\label{eq:comm_btilde}
\end{align}
the ground state with respect to $\tilde{b}_\alpha$ is defined as
$
	\tilde{b}_\alpha  |\,\Gamma\gamma\!>_0 \,=0.
$
In the subspace of the modified ground spin-states $|\,\Gamma\gamma\!>_0$,
\begin{equation}
  \mathcal{H'} = {\sum}_{\alpha}
      - 2{J'_\alpha}^2  { f_{\alpha} }^2  / {m\omega_\alpha}^2 
      +  \hbar\omega_\alpha / 2 ,
			\label{eq:HpE}
\end{equation}
where ${f_{\alpha} }^2$ 's are proportional to the unit matrix because the Pauli matrices are squared, then, the degeneracy is not lifted.
The changes in energy $\delta E'$ for the triangle and the tetrahedron are, apart from $\hbar\omega_\alpha / 2$,  
$
	-3{J'_{\rm E'}}^2 / 2 m \omega_{\rm E'}^2
$
and
$
     -  3 {J'_{\rm E}}^2 / m \omega_{\rm{E}}^2,
$
respectively, which are twice that by the static model, Eqs.~(\ref{eqkanaTE}) and (\ref{eq:statQ}).

In the subspace of the modified ground spin-states,
\begin{align}
	(Q_1,\ Q_2) = \sqrt{3/2}J'_{\rm E'} / m{\omega_{\rm E'}^2 } \, ( \sigma _x, \  \sigma_z),
		\label{eqkanaRA} 
\end{align}
for the triangle and 
\begin{align}
  (Q_{\rm u}, \ Q_{\rm v}) =-\sqrt{3}J'_{\rm E} / m {\omega_{\rm E} }^2 \, (\sigma _z, \ 
  \sigma _x)
		\label{eq:QE}
\end{align}
for the tetrahedron.
The expected values of $Q_1$ and  $Q_{\rm v}$ vanish by $\sigma_x$, 
and their fluctuations are estimated at
\begin{align}
	<{\rm{E'}}i  | {Q_1} ^2 |{\rm{E'}}i>_0  
	= 3{J'_{\rm E'}}^2/2m^2 {\omega_{\rm E'}}^4 +\hbar/2m\omega_{\rm E'}
\end{align}
with $i$=1, 2 for the triangle and
\begin{align}
  <\! {\rm E}\eta | {Q_{\rm v}} ^2 | {\rm E}\eta \!\!>_0 
  = 3{ J'_{\rm E} }^2 /m^2 {\omega_{\rm E}}^4 + \hbar/2m\omega_{\rm E},
		\label{eq:Qv2}
\end{align}
with $\eta$ = u, v for the tetrahedron.
Because the signs of the expected values of $Q_2$ and $Q_{\rm u}$ depend on the spin-states by $\sigma_z$ in Eqs.~(\ref{eqkanaRA}) and (\ref{eq:QE}),   
the clusters distort into different shapes depending on the spin-states although they remain degenerate. 
The expected values of the squared displacement $Q_1^2+ Q_2^2$ and $Q_{\rm u}^2+ Q_{\rm v}^2$ are, respectively,
$
3{J'_{\rm E'}}^2/m^2 \omega_{\rm E'}^4+\hbar/m\omega_{\rm E'}
$
and
$
 6{ J'_{\rm E} }^2 / m^2 {\omega_{\rm E}}^4   +\hbar/m\omega_{\rm E},
$
which are twice that by the static model due to quantum fluctuation, apart from the zero-point term.
Then, the change in energy by the dynamical model is twice that by static model by the virial theorem. 
On the other hand, the sum of squared expectation values is
equal to that by the static model, Eqs.~(\ref{eqkanaTE}) and (\ref{eq:statQ}). 

%
%
Let us consider about a mechanism for the tetragonal distortion at the phase transition without AF ordering in vanadium and nickel spinels with spin-1~\cite{yueda,mamiya,crawford}.
Yamashita and Ueda~\cite{YamashitaUeda00} have considered the mechanism by breaking up the tetrahedron of spin-1 composing  pyrochlore structure into the tetramer of spin-1/2. They have explained the distortion as a result of the spin driven Jahn-Teller effect.
By the present dynamical model, the distortion of $Q_{\rm u}$ component emerges because $Q_{\rm v}$ component is smeared out by the quantum-mechanical fluctuation.   
The tetragonal distortion at the structural phase transition implies the  occurrence of  hidden ordering of the nonmagnetic spin-state $| {\rm E}\eta \!\!>_0$, $\eta$ = u or v, 
because $<{\rm E}\eta | Q_{\rm u}| {\rm E}\eta \!\!>_0 $ for $\eta$ = u or v has opposite sign to each other.

\end{document}